\documentclass[12pt]{article}
\usepackage[usenames]{color}
\usepackage{fullpage}
\usepackage{graphicx}
\usepackage{graphics}
\usepackage{epsfig}
\usepackage{epsf}
\usepackage{amsmath}
\usepackage{amssymb}

\begin{document}

\title{Zero-range process with long-range interactions at a T-junction}
\author{A G Angel, B Schmittmann and R K P Zia\\
Center for Stochastic Processes in Science and Engineering,\\ 
Department of Physics, Virginia Tech, Blacksburg, VA 24061-0435, USA}
%\ead{aangel@phys.vt.edu}

\maketitle

\abstract{} 
A generalized zero-range process with a limited number of
long-range interactions is studied as an example of a transport process in
which particles at a T-junction make a choice of which branch to take based
on traffic levels on each branch. The system is analysed with a
self-consistent mean-field approximation which allows phase diagrams to be
constructed. Agreement between the analysis and simulations is found to be
very good.
%\end{abstract}
%\pacs{89.75.-k, 05.70.Ln, 05.40.-a, 05.60.-k}

\section{Introduction}

The zero-range process (ZRP) \cite{Spitzer70} is a simple model, in which
particles hop from site to site on a lattice, that has a soluble
nonequilibrium steady state for a number of cases. As such it has been
employed extensively as a model for the analysis of nonequilibrium phenomena.
Studies have ranged from fundamental investigations of nonequilibrium steady states
and phase transitions \cite{EH05,Evans00,OEC98,Godreche03,GSS03} to simple
models of real systems such as gel electrophoresis, sandpile dynamics,
traffic and compartmentalized granular gases \cite{LK92,CGS93,KMH05,Torok04}%
. Some interesting properties of the ZRP are the following. 
It has a soluble nonequilibrium steady state, as stated
before, and the statistical weight of configurations in the steady state is
described by a product measure. The relative simplicity of this product
measure form makes the steady state highly amenable to analysis. The ZRP
also displays condensation transitions where a finite fraction of the
particles in the system condense onto a single site. Of particular interest
is the fact that these transitions can take place on a one-dimensional
lattice, something that would not be expected for an equilibrium system
without long-range interactions. In homogeneous systems these phase
transitions can be of a spontaneous symmetry breaking nature, something that
has been exploited in traffic models as a possible mechanism for the `jam
from nowhere' phenomenon \cite{CSS00,Helbing01}. The ZRP has also been
proposed as a generic model for domain dynamics of one-dimensional driven
diffusive systems with conserved density, through which it provides a
criterion for the existence of phase separation in such 
models \cite{KLMST02}.

Recently, several generalizations of the basic ZRP have attracted much
attention in the literature, from both fundamental and application
standpoints. One such generalization is a system with open boundary conditions,
which can display condensation in cases where none exists for the periodic
ZRP \cite{EMS05}. Systems with multiple species of particles have been
proposed, to study inter-species mechanisms which can lead to new
condensation types \cite{EH03,HE04,Schutz03,GS03} and to model weighted directed
networks \cite{AHE05}. The ZRP has also been generalized to continuous
masses, and a general criterion for phase transitions could be stated 
\cite{ZEM04,EMZ04,MEZ05,EMZ06b}. One aspect that has long been known is the fact
that the ZRP can be solved on an arbitrary lattice \cite{Spitzer70}, i.e., with
a prescribed set of probabilities for a particle to hop from one particular
site to another. This has been generalized recently to continuous
masses \cite{EMZ06}.  ZRP's on complex networks have also attracted much  
attention recently \cite{NSL05,Noh05,TLZ06,WBBJ07,BBJW07}. 
For an overview of the ZRP and many of
its generalizations see \cite{EH05}.

The aim of this work is to investigate the ZRP with limited
long-range interactions. Specifically, we study a ZRP on
a ring lattice, a section of which consists of two branches, with
T-junctions at both ends. The rates for a particle at the junction to take
one branch or the other are based on the total numbers of particles on the
branches.  Another way of looking at this is a ZRP on an
evolving lattice, with the evolution of the lattice dependent on the state
of the system. While the connectivity of the sites is fixed, the probability of
particles taking a particular branch at the junction changes dynamically
with the state of the system. Thus, the structure of the lattice is static,
but the properties of some of the links are allowed to change. Many real
systems involve processes which take place on evolving lattices or networks.
Food webs constitute one example -- see, e.g., \cite{QHM05} -- with the
network reflecting predator-prey or "who eats whom" relations. Clearly these
interactions will change with time. For example if one species is near
extinction, its predators will preferentially prey on other species, or even
begin preying on new species entirely. Another example is a traffic
network; here the structure of the network is largely fixed, but the route
which a driver is inclined to take may change depending on the
levels of traffic in various parts of the system and any traffic calming
measures that are employed (e.g., lanes reserved for buses only during
peak times).

Often problems posed on evolving networks, possibly with long-range
interactions, are difficult to solve. Thus, it will be helpful to 
identify simple exactly or approximately \emph{soluble} models on such evolving
networks or lattices. They can contribute important steps towards the 
understanding of such systems.

The paper is organized as follows: In section~\ref{model}, the model studied
is introduced; in section~\ref{analysis} the model is analysed using the
solution of the ZRP on a fixed arbitrary lattice as the starting point for a
mean-field treatment; extensive numerical results are presented in section~%
\ref{numerics}; finally, in section~\ref{conclusion}, the implications of
this work are discussed and conclusions are drawn.

\section{The Model}

\label{model} 
\begin{figure}[tbp]
\begin{center}
\includegraphics[width=0.8\textwidth,angle=0]{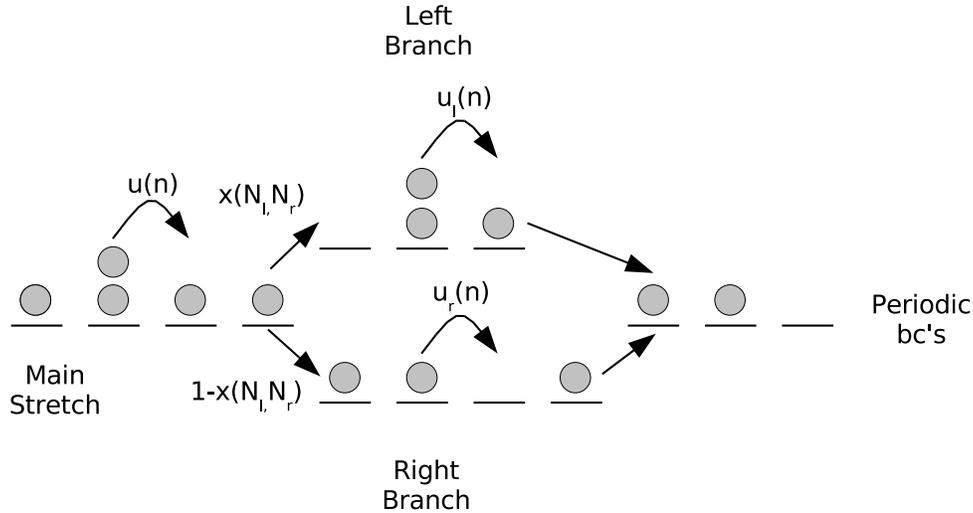}
\end{center}
\caption{Diagram of the basic system. The lattice consists of a main stretch
which splits into two branches (denoted left and right) at a T-junction
before later rejoining. The branches may have different hop rates 
and different lengths. At the first junction particles hop to the left
or right branches with probabilities that depend on the total numbers of
particles in each branch.}
\label{branchzrpdiag}
\end{figure}

The simple model studied in this paper is defined as follows.
Particles hop from site to site on a lattice which consists of a ``main
stretch'' of $L_m$ sites, branching at a T-junction into two other lanes:
labeled ``left'' and ``right'' branches, with $L_\ell $ and $L_r$ sites
respectively. At the end, these branches converge to rejoin the main stretch
(Figure~\ref{branchzrpdiag}). Since the system is periodic, the total number
of particles on the lattice, $N_{\mathrm{tot}}$, is fixed. There is no
exclusion, so that each site can hold any number of particles.  The state
of the system is completely described by the set of site occupancies 
$\{n_i\} $. Particles hop from a site $i$ with a rate $u_i(n_i)$, i.e.,
 a rate dependent on the number of particles at that site, $n_i$. This
corresponds to a particle on site $i$ hopping with a probability 
$u_i(n_i)\mathrm{d}t$ 
in a suitably small time interval, $\mathrm{d}t$. Particles hop
to the rightmost adjacent site, except at the T-junction where a 
particle hops to the left branch with a probability $x$ and to the right
branch with a probability $1-x$. If $x$ is a constant, then this system can
be solved using results for the ZRP on an arbitrary network \cite
{Spitzer70,Evans00}. In this work, we extend the model to cases where $x$
depends on the number of particles in the left branch $N_\ell $ and/or
the number of particles in the right branch $N_r$. Thus, $x$ changes
dynamically with the configuration of the system.

To keep our model as simple as possible without being completely trivial, 
we take rates that are uniform within each section and proportional to 
the standard ZRP form: $u(n)=1+b/n$, with $b>2$ so as to produce
condensation in the ordinary case \cite{Evans00,OEC98}. Specifically, we
choose, for the main stretch, left branch and right branch, respectively
\begin{subequations}
\label{hoprates}
\begin{align}
u_m(n) &= u(n) \\
u_{\ell}(n) & = p\,u(n) \\
u_r(n) & = q\,u(n) \;,
\end{align}
\end{subequations}
where $p$ and $q$ are constants. The most general scenario involves arbitrary
(positive) $p,q$. In this work, we restrict our simulations and analysis to
the unit square in the $p$-$q$ plane. Our main aim 
is to investigate what happens when a particle at
the T-junction has advance knowledge of the states of the branches and
stochastically selects which branch to take on the basis of this knowledge.
An obvious case to consider is some repulsion from highly occupied
branches. In particular, one can consider the case in which where a particle 
only has knowledge of the state of one branch. It turns out that this
situation is one of the simplest to treat analytically. An example of such a system 
is a car faced with the choice between a direct major route with frequent 
traffic reporting or a longer minor route without any traffic reporting. 
Heavy traffic may occur on either route and the question is: would it be 
faster to continue into a known jam
or take a longer route with an unknown state? A possible form of $x$ to
represent this idea is a linear dependence on the occupation of the left
branch 
\begin{equation}
x=\left[ 1-\frac{QN_\ell }{N_{\mathrm{tot}}}\right] \Theta \!\left(
1-\frac{QN_\ell }{N_{\mathrm{tot}}} \right) \;,  \label{xleft}
\end{equation}
 where $Q$ is a constant and the Heaviside function, $\Theta (n)$,
ensures that $x$, a probability, does not become negative. Another
natural case to consider is the particle making a choice based
on knowledge of the occupations of both branches. A form of $x$ that falls within this category is 
\begin{equation}
x=\frac{N_r}{N_\ell +N_r}\;.  \label{xrep}
\end{equation}
This form implies that a particle is more likely to take the branch with the
lower occupation, a realistic choice for a driver trying to avoid a jam.
Obviously, this choice embodies an extra symmetry, allowing certain
simplification in the analysis.

\section{Analysis}

\label{analysis} The simple model defined above is amenable to an
approximate analysis based on the solution of the system with a fixed $x$,
the steady state of which is known exactly 
\cite{Spitzer70,Evans00,Andjel82}.  As a precursor to the analysis for the
variable-$x$ system, results for the fixed-$x$ system are reviewed briefly.

\subsection{Fixed-$x$ system}

Focusing only on the transitions between ``fluid'' (homogeneous) and
condensed states, we may exploit the grand-canonical formalism. 
As in textbook treatments, the number of particles in such an approach is
allowed to vary and only the mean number of particles is controlled by 
the fugacity parameter: $z$. Within this framework, the steady-state distribution 
of the occupations of the sites of the system is found to be 
\begin{equation}
\begin{split}
\mathrm{P}(\left\{ n_i\right\} )& =\frac 1Z\left[ \prod_{i\in \mathrm{ms}%
}f_m(n_i)z^{n_i}\right] \left[ \prod_{j\in \mathrm{lb}}f_\ell
(n_j)(zx)^{n_j}\right] \\
& \quad \quad \quad \quad \left[ \prod_{k\in \mathrm{rb}%
}f_r(n_k)(z(1-x))^{n_k}\right]
\end{split}
\;,
\end{equation}
where the products go over the sites which are in the main stretch (ms),
the left branch (lb) and the right branch (rb), respectively. Here, 
\begin{equation}
f_\mu (n)=\left\{ 
\begin{array}{cc}
\displaystyle{\prod_{k=1}^n}u_\mu (k)^{-1}\quad & \mbox{for $n>0$} \\ 
1\quad & \mbox{for $n=0$}
\end{array}
\right. \;,
\end{equation}
and obviously, the subscript $\mu = m$, $\ell$, $r$ for the ms, lb, and rb, 
respectively. Finally, $Z$ is a
normalization constant, akin to the grand-canonical partition function, 
\begin{equation}
\begin{split}
Z=\sum_{n_1=0}^\infty \sum_{n_2=0}^\infty \cdots \sum_{n_L=0}^\infty &
\left[ \prod_{i\in \mathrm{ms}}f_m(n_i)z^{n_i}\right] \left[ \prod_{j\in 
\mathrm{lb}}f_\ell (n_j)(zx)^{n_j}\right] \\
& \qquad \left[ \prod_{k\in \mathrm{rb}}f_r(n_k)(z(1-x))^{n_k}\right]
\end{split}
\;.
\end{equation}
With the chosen hopping rates (\ref{hoprates}), 
we have 
$f_m(n)=f(n)$, $f_\ell (n)=p^{-n}f(n)$ and $f_r(n)=q^{-n}f(n)$,
so that this expression simplifies considerably:
\begin{equation}
Z=\sum_{ \left\{ n \right\} }
\left[ \prod_{i\in \mathrm{ms}}f(n_i)z^{n_i}\right] 
\left[ \prod_{j\in \mathrm{lb}}f(n_j)(zx/p)^{n_j}\right]
\left[ \prod_{k\in \mathrm{rb}}f(n_k)(z(1-x)/q)^{n_k}\right]
\;.
\end{equation}

\subsubsection{Condensation transitions in the fixed-$x$ system}

Let us first remind the reader that the grand-canonical treatment
gives the correct distribution of the system only below a certain critical
density $\rho _{\mathrm{c}}$. Of course, due to the presence 
of the branches, this quantity will depend on the parameters $(x,p,q)$. 
To see how this arises, we will need the expression which relates the 
overall density of the system ($\rho \equiv N_{\mathrm{tot}} /L$)
to the fugacity $z$. But we have three sections and the average density of 
each must be considered separately. To facilitate, let us define
\begin{equation}
g \left(  \zeta \right) \equiv \frac{\sum_{n=0}^{\infty} n  \zeta ^n 
f(n)}{\sum_{n=0}^{\infty}  \zeta ^n f(n)} \;,
\end{equation}
which can be used to relate the density in each section to the effective fugacities: 
\begin{equation}
z_m \equiv z ; \qquad z_{\ell} \equiv zx/p ; \qquad z_r \equiv z(1-x)/q
\;.
\end{equation}
Summarizing, we write
\begin{equation}
\rho_{\mu} \equiv \left< N_{\mu} \right> / L_{\mu} = g\left( z_{\mu} \right)
\;,
\end{equation}
so that 
\begin{equation}
\begin{split}
\rho & = \left< \frac{N_m+N_\ell+N_r}L \right> = 
\sum_{\mu} \rho_{\mu} \frac{L_{\mu}}L \\
& =\frac{L_m}L g\left( z \right)
+\frac{L_\ell }L g\left( zx/p \right)
+\frac{L_r}L g\left( z(1-x)/q \right) \;.
\end{split}
\label{densityeqn}
\end{equation}

From our experience with standard ZRP's, it is clear that this approach 
is valid as long as each effective fugacity remains less than unity (so
that $Z$ remains finite). Thus, $z$ cannot exceed the minimum of $1$, $p/x$, 
$q/(1-x)$. Meanwhile, $g$ is a monotonically increasing function of its 
argument and, if $g(1)$ is finite, then there will be a density, 
$\rho _{\mathrm{c}}$, beyond which (\ref{densityeqn}) has no solution.
This happens when the hop-rates in each section decay to some constant value 
$\beta $ more slowly than $\beta (1+2/n)$ \cite{Evans00,OEC98}. In the
system, the lack of a solution to the density equation manifests itself as a
symmetry-breaking condensation transition whereby the excess density will be
taken up by a single site. In our case, this site will be located in the
section where the effective fugacity first reaches unity. For times much
longer than the characteristic time scale associated with individual 
particles hopping, the condensate will be found on
just one site in this section. In simulations starting from random initial
conditions, we often observe the condensate to form on the first site of
a stretch. However, with other initial conditions the condensate can
form anywhere and we believe that, for finite systems and long
times, the condensate will move slowly between sites, eventually
exploring all of the possible locations. 

To summarize, if $u(n)$ decays more slowly than $1+2/n$, then we can expect 
condensation for high densities. Further, the condensate will appear in the
main stretch if $p>x$ and $q>1-x$. Otherwise, it will appear in the left or 
the right branch, depending on whether $p(1-x)$ is less or greater than
$qx$. Though the critical density needed for each of the sections is the 
same (i.e., $g(1)$), the overall critical density will depend on the details 
of the parameter set. As an illustration, suppose $x=0.3$, $p=0.2$, and $q=0.8$, 
so that $z$ should not exceed $2/3$. A condensate will then appear on the left 
branch when the overall density exceeds 
$\left[ L_m g(2/3)+ L_\ell g(1)+ L_r g(7/12) \right] / L$.

\subsection{Varying-$x$ system}

The aim is now to use the exact solution of the fixed-$x$ system as a tool
to understand the system with the dynamically changing branch choice. The
basic assumptions is that if the system relaxes to a steady state, it will do so with
some well-defined average value for $x$. In particular, this assumption is valid
if the fluctuations of the particle numbers entering the definition of 
$x$, (\ref{xleft}) or (\ref{xrep}), are not too large.  
%Given that even in the fixed-$x$
%system $x$ is just a probability, it is reasonable to expect that if the
%fluctuations in the varying $x$ are not too large, the system will take on a
%very similar distribution to the fixed-$x$ system with the average value of 
%$x$ used as the fixed value. 
Within this self-consistent
mean-field approximation the average numbers of
particles in the left and right branches (with the exception of any
condensate) can be calculated as a function of $x$:
\begin{equation}
\left< N_{\ell} \right> = L_{\ell} g(zx/p)\;, \qquad 
\left< N_r \right> = L_r g(z(1-x)/q)\;. 
\end{equation}
For our specific hopping rates, $g$ is explicitly 
\begin{equation}
g\left( \zeta \right) = \frac{\zeta \,{}_2\mathrm{F}_1(2,2;2+b;\zeta )}%
{(1+b)\,{}_2\mathrm{F}_1(1,1;1+b;\zeta )}\;
\end{equation}
where ${}_2\mathrm{F}_1$ is the hypergeometric function 
\cite{specfunbook}. Note that $g(1)=1/(b-2)$, which is finite for $b>2$ and 
can provide the quantitative aspects of the phase boundaries. The expressions 
for $\left< N_{\ell} \right>$ and $\left< N_r \right>$ can then be fed into a 
self-consistent equation for $x$. Whether or not this equation admits a solution 
reveals much about the system.
Although the grand-canonical treatment can describe only the sub-critical
behaviour of the system in detail, it does lead us to some information 
about the phase diagram. As we will see, it can predict some simple aspects of
the condensed phases. For example, for the fixed-$x$ system, a single
condensate appears to soak up all the excess mass in the system. However, for
the varying-$x$ system, the feedback mechanism seems to be able to prevent a
single condensate from absorbing all of the excess mass. Instead, the
excess is shared between two or more condensates.

\subsubsection{Branch choice dependent on left branch only}

The first case chosen for study has the branch choice probability, $x$,
dependent on the state of the left branch only. For this system, which is
defined by the branch choice probability (\ref{xleft}), the self-consistent
equation for $x$ is 
\begin{equation}
x=1-\frac{QL_\ell }{N_{\mathrm{tot}}}g\left( zx/p\right)
\label{lbxconsistency}
\end{equation}
This equation depends on both $x$ and $z$, which are also partially
dependent on each other. Within the self-consistent scheme we
proposed, both (\ref{lbxconsistency}) and (\ref{densityeqn}) are to be
solved simultaneously in $x$ and $z$. If a simple solution exists, this
implies that there is no condensation. Much like the case for Bose-Einstein
condensation presented in textbooks, if the naive approach fails to produce a
solution, the specifics of this failure will provide enough
information for us to predict where condensation occurs.

Recall that in the fixed-$x$ system the maximum allowed value of the
fugacity, $z$, was $z_{\mathrm{max}}=\min \{ 1,p/x,q/(1-x) \}$. Thus if (\ref
{lbxconsistency}) can be solved along with (\ref{densityeqn}) under this
constraint on $z$, then there will be no condensation and the number of
particles in each section will be known simply by inserting the value of $x$
from the solution into the relevant expression.

If, under the constraints on $z$, (\ref{lbxconsistency}) can be solved and (%
\ref{densityeqn}) cannot, then this implies that a condensate exists but not
on the left branch. Recall that in the fixed-$x$ system, inserting $z$ at
its maximum value into any of the expressions above provides a correct
description of the system except for the condensed site. Thus, (\ref
{lbxconsistency}) is solved for the maximum value of $z$. The location of
the condensate can then be determined by inspecting $z$ and $z(1-x)/q$. If $%
z<z(1-x)/q$, then the condensate appears on the right branch, otherwise it
appears on the main stretch.

If neither (\ref{lbxconsistency}) nor (\ref{densityeqn}) can be solved, 
this suggests that $x$ takes a value such that there is a condensation on
the left branch. However, this condensation may not be the same as in the
fixed-$x$ system, since the form (\ref{xleft}) tends to suppress
large numbers of particles on the left branch. In the fixed-$x$ system the
condensate grows indefinitely with increasing total particle density.
However, in this varying-$x$ system, an increase of the number of particles on
the left branch results in a decreasing value of $x$. This implies that a
condensate on the left branch will not be able to take up the excess density
for all densities. In this case, the maximum value of $z$ will be such that
either $zx/p=z>z(1-x)/q$ or $zx/p=z(1-x)/q>z$. That is, the system organizes
itself into such a state that two coexisting condensates are supported:
In the grand-canonical ensemble two sections are critical and both hold
a condensate. The locations of the coexisting condensates can be determined
from which of $z$, $zx/p$ and $z(1-x)/q$ are equal, e.g., if $z=zx/p$ then
the condensates appear on the left branch and the main stretch.

Unfortunately, the self-consistent and density equations (\ref
{lbxconsistency}), (\ref{densityeqn}) are often difficult to solve
analytically, due to the presence of hypergeometric functions.
However, they can be solved numerically and this can be used to map out the
phase diagram of the system in the parameters $p$ and $q$.

The phase diagram becomes particularly easy to calculate if the following
two assumptions are made. The number of particles in the system is large
enough that the density equation (\ref{densityeqn}) cannot be solved for any
allowed values of $z$ and $x$. Thus there must always be a condensate
somewhere in the system. This allows one to assume that $z$ must be at its
maximum value of $\min \left[ 1,p/x,q/(1-x)\right] $, instead of numerically
solving the density equation. Also, $Q$ is sufficiently large that a
condensate cannot appear on the left branch only. The phase diagram of this
system can then be mapped out by first assuming a phase, and then working out
the values of $p$ and $q$ necessary for its presence.

The phases are identified by the location(s) of the condensate(s) in the
system: L for the left branch, R for the right branch and M for the main
stretch, while N indicates that no condensates are present in the system
Thus, LR denotes a phase where condensates appear on the
left and right branches, for example.

\textbf{M Phase---}For a condensate to be present only on the main stretch, 
$z$ must be equal to $1$, both $p/x$ and $q/(1-x)$ must be greater than $1$
and the self-consistent equation (\ref{lbxconsistency}) for $x$ must have a
solution. Thus, this phase is bordered by the line $p=\min (x)$, where $\min
(x)$ refers to the smallest possible value of $x$ that can be found for
(\ref{lbxconsistency}) in the grand-canonical treatment. It is given 
by 
\begin{equation}
\min(x)=1-\frac{QL_{\ell}}{N_{\mathrm{tot}}(b-2)}\;. \label{lbxminx}
\end{equation}
It is also
bordered by a line which comes from the solution of the self-consistent equation
(\ref{lbxconsistency}) where $q/(1-x)$ becomes less than one indicating a
shift to an R phase.  This line is given by 
\begin{equation}
q=\frac{QL_{\ell}g(r)}{N_{\mathrm{tot}}} \;, \label{MRborder}
\end{equation}
 where $r=(1-q)/p$.

\textbf{LM Phase---}For condensates situated on the left branch and the main
stretch to coexist, $z$ must be equal to $1$, $p/x$ must be equal to $1$ (so 
$x=p$), $q/(1-x)$ must be greater than $1$ and the self-consistent equation (%
\ref{lbxconsistency}) must have no solution. Thus, this phase is bordered by
a line already identified for the M phase ($p=\min (x)$) and also by the
line $q=1-p$.

\textbf{LR Phase---}For condensates situated on the left and right branches
to coexist, $z$ must be less than $1$, $p/x$ must be equal to $q/(1-x)$,
both of these must be equal to $z$ and the self-consistent equation for $x$ (%
\ref{lbxconsistency}) must not have a solution. Thus, this phase is bordered
by a line already identified for the LM phase ($q=1-p$) and the line $%
q=(1/\min (x)-1)p$ which is the border for (\ref{lbxconsistency}) to have a
solution.

\textbf{R Phase---}For a condensate to be present on the right branch only, $%
z$ must be less than $1$, $p/x$ must be greater than $z$ and $q/(1-x)$ must
be equal to $z$. The boundaries for this phase have already been
identified as the boundaries of two other phases; this phase is bounded by
the line $q=(1/\min(x) -1)p$ and the line which comes from the solution of
the self-consistency equation (\ref{lbxconsistency}) where $x$ is such that $%
q/(1-x) = 1$ which is given by (\ref{MRborder}).

The other possible phases (L, RM, LRM and N) are not observed when both the
density of particles and the repulsion parameter, $Q$, are sufficiently high,
as has been assumed above. For an L phase the density must be sufficiently
low, or the repulsion sufficiently weak, to hold a condensate without $x$
becoming too small to sustain it. For an RM phase to be present, there must
be some repulsion from the right branch, otherwise the condensate will
prefer to form on only one of the sections. For an LRM phase one must have 
$z=1$, $zx/p=1$ and $z(1-x)/q=1$, which can only happen on the line $q=1-p$, 
and even so, it is likely that at
least one of the sections will merely be at criticality and not hold a
condensate. Finally for the N phase the density must be so low
that there does not need to be a condensate anywhere in the system.

\begin{figure}[tbp]
\begin{center}
\includegraphics[width=\textwidth,angle=0]{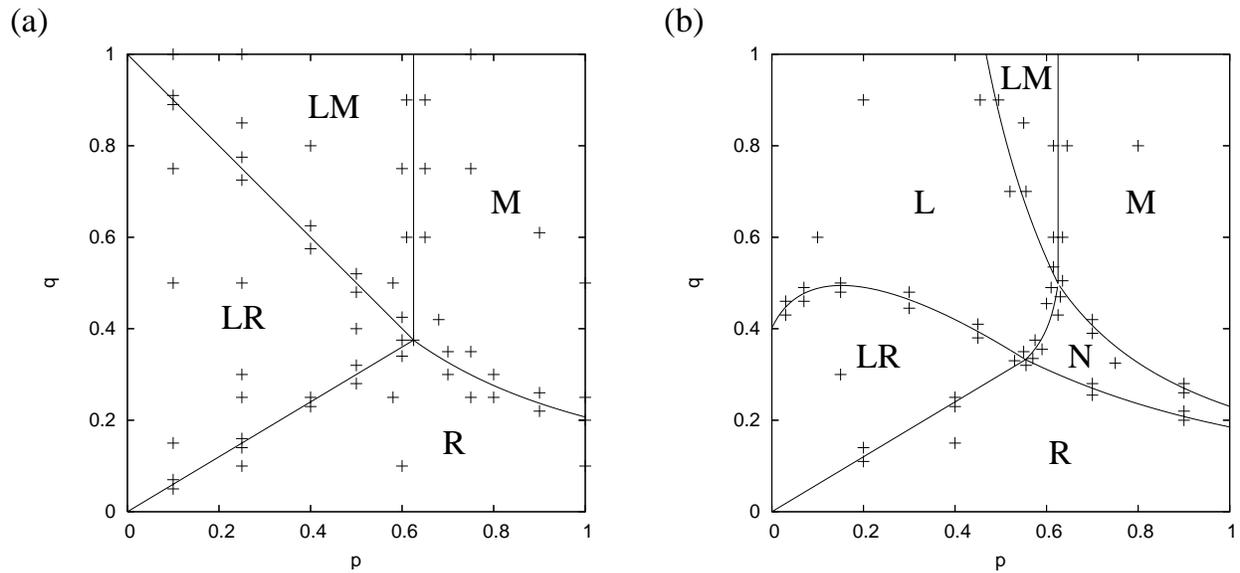}
\end{center}
\caption{Phase diagrams for the left-branch feedback system with $L_{\ell} =
300$, $L_r = 600$, $L=1900$ and (a) $Q=20$, $N_{\mathrm{tot}} = 8000$ and
(b) $Q=2$, $N_{\mathrm{tot}}=800$. Also shown are points where simulations
have been run and the expected behaviour verified. The phases are labelled
by the position(s) of the condensate(s): L indicates a condensate on the
left-branch, R indicates a condensate on the right-branch, M indicates a
condensate on the main stretch and N indicates that no condensates are
present.}
\label{phasediag}
\end{figure}

Assembling all this information together, phase diagrams for given
parameters can be produced. As an example, the phase diagram for 
$Q=20$, $N_{\mathrm{tot}}=8000$, $L_\ell =300$, $L_r=600$ and $L_m=1000$
calculated via this method is shown in Figure~\ref{phasediag} (a). Also
shown are some of the values of $p$ and $q$ for which simulations were run
to verify the phase behaviour.

It is straightforward to calculate how the phase diagram will be
affected by changing the repulsion parameter $Q$ and/or the lengths of the
branches. The fact that all the phase boundaries meet at a single point does
not change when these parameters are varied; this point is simply displaced.
Assuming a sufficiently high density, the line $q=1-p$ always forms the
boundary between the LM and LR phases; only its length depends on the
parameters noted above. The vertical boundary between the LM and M phases is
displaced horizontally and its length changes also. For the boundary between
the LR and R phases, the slope changes. Finally the boundary between the R
and M phases retains its curved nature, but is displaced and
shortened/lengthened.

The theory can also be applied to systems at lower densities and with weaker
repulsion, but it is less straightforward. The simplification associated
with high density systems - $z$ being pinned to a maximum value - no longer
holds. As a result, we must generally rely on numerical methods to find a
solution for both equations (\ref{lbxconsistency}) and (\ref{densityeqn}). With
this approach, the phase boundaries can be mapped out, as in the specific
case of $Q=2$, $N_{\mathrm{tot}}=800$, $L_\ell =300$, $L_r=600$ and $%
L_m=1000 $ (Figure~\ref{phasediag} b). For these parameters, the N and L
phases are realised. We also see that the phase diagram has a much richer
structure.

A simpler task than computing the full phase diagram for any particle number
is calculating how and when the phases emerge.  Clearly at low enough particle
number, no condensates will be present.  The M phase will emerge at $p=q=1$
when $N_{\mathrm{tot}}=L_{\ell}g(x) + L_r g(1-x) + L_m g(1)$, with $x$ coming from the
solution of $x=1-Q L_{\ell} g(x)/(L_{\ell}g(x) + L_r g(1-x) + L_m g(1))$.  For $Q=2$ this takes place at
$N_{\mathrm{tot}}=610$.  The L (R) phase first emerges at sufficiently small $p$ ($q$)
when there are enough particles to support such a phase, i.e.,
$N_{\mathrm{tot}}=L_{\ell} g(1)$ ($L_r g(1)$) and the LR phase when there are enough
particles to support both condensates, $N_{\mathrm{tot}}=(L_{\ell} + L_{r}) g(1)$ and
both $p$ and $q$ are sufficiently small.  The LM phase will emerge from the
$q=1$ boundary when both the L and M phase can first support condensates at
$N_{\mathrm{tot}}=L_{\ell} g(1) + L_r g(1-x) + L_m g(1)$ where $x$ is given by
$x=1-(Q/N_{\mathrm{tot}}) L_{\ell}g(1)$; for $Q=2$ this is $N_{\mathrm{tot}}=714$.  
The R and M phases first meet at the $p=1$ boundary when
$N_{\mathrm{tot}}=L_{\ell}g(1-q) + (L_r + L_m)g(1)$, with $q$ coming from the solution
of $q = Q L_{\ell} g(1-q) / (L_{\ell}g(1-q) + (L_m+L_r)g(1))$, which in the
case $Q=2$ corresponds to $N_{\mathrm{tot}}=869$.  The N phase disappears when the LR
and LM phases meet up at $N_{\mathrm{tot}}=(L_{\ell}+L_r+L_m)g(1)$, the first point
where all three sections can have critical occupancies.  The LR and LM phases
then begin to border each other along the line $q=1-p$ and this line extends 
as $N_{\mathrm{tot}}$ is increased until the L phase is extinguished at $N_{\mathrm{tot}}=Q(L_r
+ L_m)g(1)/(Q-1)$.

\subsubsection{Branch choice dependent on left and right branches}

Next, we study the case where the branch choice probability depends
on the occupation of both the left and right branches in a repulsive
fashion, i.e., the value of $x$ favours sending particles to the
least occupied branch. The form used, given in (\ref{xrep}), is 
clearly more symmetric than the previous case. We can write the
self-consistent equation for $x$ in a way that makes this symmetry transparent: 
\begin{equation}
L_r (1-x) g\left( z(1-x)/q\right)=L_\ell x g\left( zx/p\right) \;.
\label{bbxconsistencyeqn}
\end{equation}
Again, the solution of this equation (or lack thereof) in
conjunction with the density equation (\ref{densityeqn}) reveals a
great deal about the system.

The phase diagram for this system can be mapped out in much the same
manner as above, though the analysis is slightly more involved.

As before, simplifications occur if the total number of particles in the
system is high enough so that a condensate must appear somewhere in the
system. Then, the phase diagram can be constructed by assuming the
location(s) of the condensate(s) and examining the values of $q$ and $p$ for
which a solution can persist. As an example, in order for the system to
display a condensate on only the main stretch, we must have $z=1$ and a
solution to the self-consistency equation (\ref{bbxconsistencyeqn}) such
that both $x<p$ and $(1-x)<q$.

For a sufficiently large $N_{\mathrm{tot}}$,
there are three possible ways in which equation (\ref{bbxconsistencyeqn})
cannot be solved. For the system to be in the L, R, or LR phase, we must
have $x=p$, $(1-x)=q$, or $x=p$ and $(1-x)=q$ respectively. By
calculating when these equalities hold in conjunction with $z$ being pinned
to its maximum value of $\min \{1,p/x,q/(1-x)\}$, the rest of the phase
diagram is mapped out.

\textbf{M Phase---}As noted above, for a condensate to be present
only on the main stretch, $z$ must be equal to $1$, both $p/x$ and $q/(1-x)$
must be greater than $1$ and the self-consistent equation 
(\ref{bbxconsistencyeqn}) for $x$ must have a solution. 
Thus, this phase is bordered by a
line which comes from the numerical solution of (\ref{bbxconsistencyeqn})
which in this case takes the form
\begin{equation}
L_r(1-x)g((1-x)/q) = L_{\ell}xg(x/p)\;. \label{Mboundary}
\end{equation}
When no solution exists, the system must shift to a phase with a
condensate present on one or more of the branches.

\textbf{L Phase---} For a condensate to exist only on the left branch, we
require $z<1$, $z=p/x$, $z<q/(1-x)$ and no solution to the self-consistent
equation (\ref{bbxconsistencyeqn}) for $x$ such that the occupation of the
left branch is greater than the critical value, but the occupation of the
right branch is not.  Thus, in the region $q<1-p$ this phase is bordered by
the line coming from the solution of
\begin{equation}
  N_{\mathrm{tot}}=L_r g(1) (p+q)/p + L_m g(p+q) \;. \label{L-LRphaseboundary}
\end{equation}
Along this line the system shifts into a phase with condensates present on the
left and right branches.  In the region $q>1-p$ this phase is bordered by the
line from the solution of 
\begin{equation}
N_{\mathrm{tot}}=L_r g((1-p)/q)/p + L_m g(1) \;.   \label{L-LMphaseboundary}
\end{equation}
Along this line the system shifts into a phase with condensates present on the
left branch and the main stretch.

\textbf{R Phase---}For a condensate to exist only on the right branch, we
require $z<1$, $z<p/x$, $z=q/(1-x)$ and the right branch occupation to be
greater than critical so that there is no solution to the self-consistent
equation for $x$ (\ref{bbxconsistencyeqn}).  Thus in the region $q<1-p$ this
phase is bordered by a line coming from the solution of 
\begin{equation}
N_{\mathrm{tot}} = L_{\ell} g(1) p/q + L_m g(p+q) \;. \label{R-LRphaseboundary}
\end{equation}
Along this line the system moves into the LR phase.  In the region 
$q>1-p$ the phase is bordered by a line given by
\begin{equation}
N_{\mathrm{tot}} = L_{\ell} g((1-q)/p) p/(1-p) + L_m g(1) \;.  \label{R-RMphaseboundary}
\end{equation}
Along this line the system moves into the LM phase.

\textbf{LM Phase---}For condensates situated on the left branch and the main
stretch to coexist, we must have $z=1$, $x=p$, $1-x<q$, and no solution
to the self-consistent equation for $x$ (\ref{bbxconsistencyeqn}). Thus,
this phase is bounded by the line already identified as a boundary for the L
phase (\ref{L-LMphaseboundary}),
the line $q=(1-p)$ and a line coming from the
numerical solution of (\ref{bbxconsistencyeqn}) which in this case takes the form 
\begin{equation}
L_r(1-p)g((1-p)/q)=L_{\ell}pg(1)\;,
\end{equation}
and gives the limit of the region in which no solution can be found. 
This coincides with part of the boundary from the M phase; in fact, the 
only way in which the equation giving the boundary for the M phase 
(\ref{Mboundary}) cannot be solved under the high density assumption 
is if either $x=p$ or $(1-x)=q$.
%\begin{equation}
%\[
%\frac 1p=1+\frac{q\,{}\mathrm{F}_2\times \,{}_2\mathrm{F}_1(1,1;1+b;(1-p)/q)%
%}{(1-p)\mathrm{F}_1\times \,{}_2\mathrm{F}_1(2,2;2+b;(1-p)/q)}
%\]
%\;.  \label{playing}
%\end{equation}
%\begin{eqnarray}
%&&
%\[
%g_1\left( \xi \right) \equiv \,{}_2\mathrm{F}_1(1,1;1+b;\xi )
%\]
%\label{defining new g's} \\
%g_2\left( \xi \right)  &\equiv &...\;.
%\end{eqnarray}

\textbf{RM Phase---}Similarly, for condensates situated on the right
branch and the main stretch to coexist, we must have $z=1$, $x<p$, $1-x=q$,
and no solution to the self-consistent equation for $x$ (\ref
{bbxconsistencyeqn}). Thus, this phase is also bounded by the line $%
q=(1-p)$, a line (different from the one above) coming from the
limits of the numerical solution of (\ref{bbxconsistencyeqn}) and a line
already identified as a boundary for the R phase (\ref{R-RMphaseboundary}). 
The first line corresponds to the part of the M boundary not already 
matched by the LM boundary
\begin{equation}
L_r q g(1) = L_{\ell}(1-q)g((1-q)/p)\;.
\end{equation}
Note that if
the lengths of the left and right branches are the same, as chosen here,
this phase is symmetric with the LM phase.

\textbf{LR Phase---}Finally, for condensates situated on the left and
right branches to coexist, the appropriate conditions are $z<1$, $x=p$, $%
1-x=q$, and no solution to the self-consistent equation for $x$ (\ref
{bbxconsistencyeqn}). Thus, this phase is bounded by the line $q=1-p$, a
line from solving (\ref{bbxconsistencyeqn}) which happens to touch, but not
cross, $q=1-p$ for the case studied here and lines that have already been
identified as boundaries for the L and R phases 
(\ref{L-LRphaseboundary}),
(\ref{R-LRphaseboundary}).

As before, the other possible phases (LRM and N) are not observed when
the density is sufficiently high, as has been assumed.  The
LRM phase can again only possibly exist on the line $q=1-p$ 
and it is likely that the condensate will only form on at most two of the 
sections. Finally the $N$
phase 
will not be seen when the density is sufficiently high as the system must
take on a condensate to accommodate the number of particles in a steady state.

\begin{figure}[tbp]
\begin{center}
\includegraphics[width=\textwidth,angle=0]{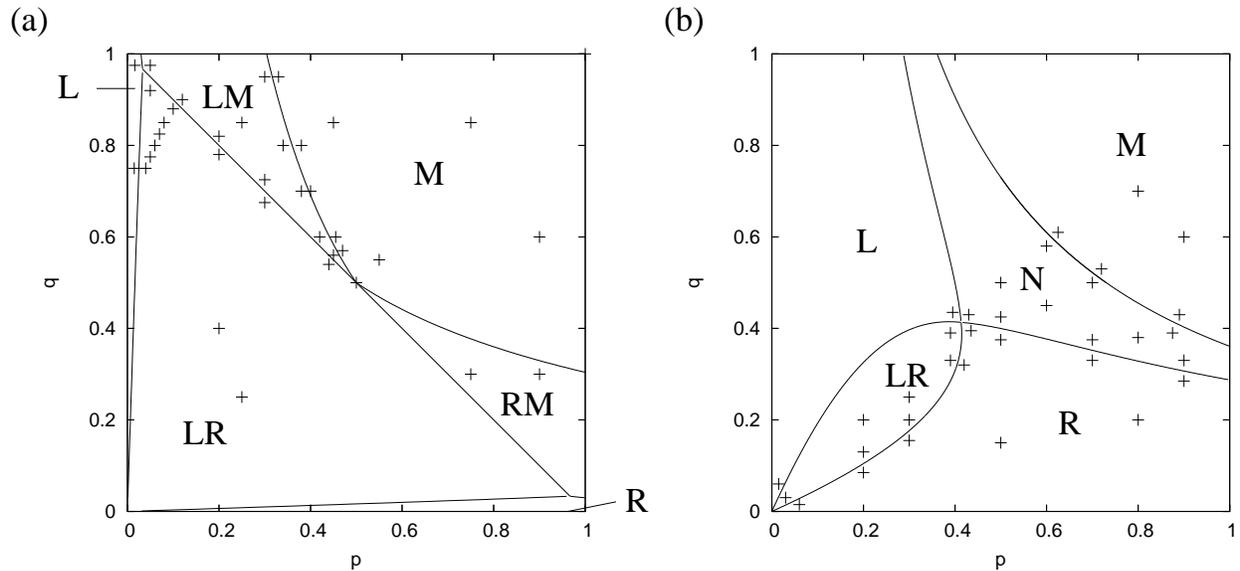}
\end{center}
\caption{Phase diagrams for the system with repulsive feedback from both
branches, with $L_{\ell} = L_r = 500$, and $L_m=1000$ and (a) $N_{\mathrm{tot}} =
8000 $, (b) $N_{\mathrm{tot}}=800$. Also shown are points where simulations have been
run that verify the predicted behaviour. Note that only half of the diagrams
have been explored in this way due to the symmetry of the system. The phases
are labelled by the position(s) of the condensate(s): L indicates a
condensate on the left branch, R indicates a condensate on the right
branch, M indicates a condensate in the main stretch and N indicates
that no condensates are present.}
\label{phasediagrep}
\end{figure}

The phase diagram for a system with $N_{\mathrm{tot}}=8000$, $L=2000$ and $%
L_{\ell} = L_r = 500$ is shown in Figure~\ref{phasediagrep} (a). Also shown
are the $p$ and $q$ values for simulations that have been run to verify the
phase behaviour.

As before the changes in the phase diagram due to changing the lengths of the
branches (which do not have to be the same) are straightforward to compute.
The boundary of the LR phase remains unchanged as the line $q=1-p$, but the
point where the other boundary lines meet with this one does move along this
line and the boundaries between the LM and L and LR and L change length and
slope accordingly. The boundary lines between the M and the LM and RM phases
respectively retain their curved shape, but they move and the points at which
they touch $q=1-p$ and $q=1$ or $p=1$ also change.

As in the previous, case, the analysis can also be applied to systems with 
densities that do not a priori guarantee the presence of a condensate. 
Again, however, the application is less straightforward and relies on simultaneous numerical
solutions of the density equation (\ref{densityeqn}) and the self-consistent
equation for $x$ (\ref{bbxconsistencyeqn}). The phase diagram calculated in
this way for the system with $L_{\ell} =L_r=500$, $L_m=1000$ and $N_{\mathrm{tot}}=800$
(i.e., the same system as above but at a lower density) is shown
in Figure~\ref{phasediagrep} (b). Here the N phase is realised, the
LM and RM phases are lost and the phase diagram has become richer in
structure.

Again, how and when the phases emerge as the total number of particles is increased
is perhaps more straightforward to calculate than the full phase diagram.
Clearly, at low enough particle numbers there will be no condensation in the
system.  For sufficiently small $q$ ($p$) the R (L) phase will emerge simply
when there are sufficient particles to support such a phase i.e., $N_{\mathrm{tot}} =
L_r g(1)$ ($L_{\ell} g(1)$).  Likewise, the LR phase will emerge for
sufficiently small $p$ and $q$ when $N_{\mathrm{tot}} = (L_l+L_r) g(1)$.  The M phase
must first emerge at $p=q=1$ when the number of particles on the main stretch
first becomes critical, and will do so at $N_{\mathrm{tot}} = L_m g(1) + L_r g(1-x) +
L_{\ell} g(x)$ with $x$ being the solution of $x( L_r g(1-x)+L_{\ell} g(x)) =
L_r g(1-x)$.  For the symmetric system with $L_m=1000$, $L_{\ell}=L_r=500$
this turns out to be $N_{\mathrm{tot}}=634$.  The RM phase emerges from the $p=1$
boundary at when both the right branch and main stretch occupancies first
attain their critical values together.  This is at the point $N_{\mathrm{tot}}=L_r
g(1)/(1-q) + L_m g(1)$ with $q$ coming from the solution of $ (1-q)(L_{\ell}
g(1-q) + L_r g(1)) = L_r g(1)$; for the specific system studied here this is
$N_{\mathrm{tot}}=860$.  Also, for the specific system studied here the RM phase is
symmetric with the LM phase.  At $N_{\mathrm{tot}}=1000$, the N phase disappears
completely as this is the point where the LR, LM and RM phases first meet up 
i.e., when $N_{\mathrm{tot}}=L_{\ell}g(1)+L_r g(1) + L_m g(1)$.  The L and R phases are
present for all high particle densities, although they will eventually become
infinitesimally small, the thickest part of the phases behaving as $p,q\sim
1/N_{\mathrm{tot}}$ respectively. 

Finally, we note that in the condensed phases the size of the condensate(s)
can change to allow a solution of both the $x$ balance equation and the
density equation.  This will break down only when the $x$,$z$ solution becomes
such that the condition for another phase will be met or when the condensate
has to take its critical value to satisfy the equations.  Due to the way that
the occupations of each section depend on $z$ it can be shown that approaching
a boundary from either side must result in an unambiguous answer.  Thus, it is
expected that many different forms of the $x$-function will give
the kind of multiple condensate behaviour seen for the two choices, 
(\ref{xleft}) and (\ref{xrep}), considered in this study. 
The only constraints which should be placed on a suitable $x$-function are 
the following: it has a range between $0$ and $1$; it gives a
unique solution to the self-consistent equation for $x$ and the density 
equation for any given ($p$,$q$) pair; it is repulsive in some way so as to inhibit a
sole condensate; and it is sufficiently smooth that the solution will not be
discontinuous within any phase.

\section{Numerical Results}
\label{numerics} To verify the predicted phase diagrams, we performed
extensive Monte Carlo simulations on this system with both feedback
mechanisms. Our simulation method is simple: A site is picked at random and
a particle there is moved to the next site with the relevant probabilities,
(\ref{hoprates}), (\ref{xleft}), and (\ref{xrep}). We have studied a range of branch
lengths, $N_{\mathrm{tot}}$'s, $p$'s, $q$'s and $Q$'s. In this paper, we only present the results
of $L_m=1000$, $0<p<1$, $0<q<1$, $N_{\mathrm{tot}}=\{800,8000\}$, with $L_{\ell}=300$, $L_r=600$, 
$Q=20$ for the 
system with feedback from the left branch only and $L_{\ell}=L_r=500$ 
for the system with feedback from both branches. In the phase diagrams, 
shown in Figures~\ref{phasediag} and \ref{phasediagrep}, we display the behaviour of the
condensates. N indicates no condensates are present. If a single
condensate appears in the system, its location is denoted by L, R, and M -
indicating the left, right, and main sections, respectively. Similarly, the
labeling of coexistence of two condensates is self-explanatory. The lines are
predictions of the phase boundaries from our mean-field theory. The crosses
(+) indicate points at which simulations have been performed. All numerical
results showed the expected behaviour for the relevant phase. Occasionally,
small discrepancies between prediction and simulation were seen very close
to the boundaries, but in all cases these discrepancies lessened when larger
system sizes were used. Note that, since systems with the second feedback
mechanism ((\ref{xrep}) and Figure~\ref{phasediagrep}) are symmetric 
under $p\leftrightarrow q$, only one half of the phase diagram 
(above or below the line $q=p$) was tested.

One phenomenon appeared in the majority of the runs, namely, a tendency for
condensates to form on the first site of the section. The reason for this
behaviour is neither transparent nor intuitive. The exact solution for the
fixed-$x$ system implies that the condensate is equally likely to be found
anywhere in the \emph{stationary} state. Clearly, the \emph{dynamics}
breaks overall translational invariance, so that the favouring of the first
site may be a subtle manifestation of the underlying dynamics. Exploring
this issue is beyond the scope of this work, but would be interesting for
future studies. To insure that other sites are equally favoured, in the
steady state, for condensation, we carried out runs where one or more
condensates were initially placed at ``interior'' sites. It is reassuring
that such condensates did not move to the first site. We should emphasize,
of course, that the probability for a condensate to appear on any site (of
the allowed sections) is equal, so that the condensate will move between
sites. Given sufficient time, all sites will eventually
be explored in a \emph{finite} system. 
This wandering behaviour has been observed, especially when
a small condensate on one section coexists with a large condensate in
another: The large condensate remains stationary, but the smaller one moves
between several sites over the duration of the run.

\begin{figure}[tbp]
\begin{center}
\includegraphics[width=\textwidth,angle=0]{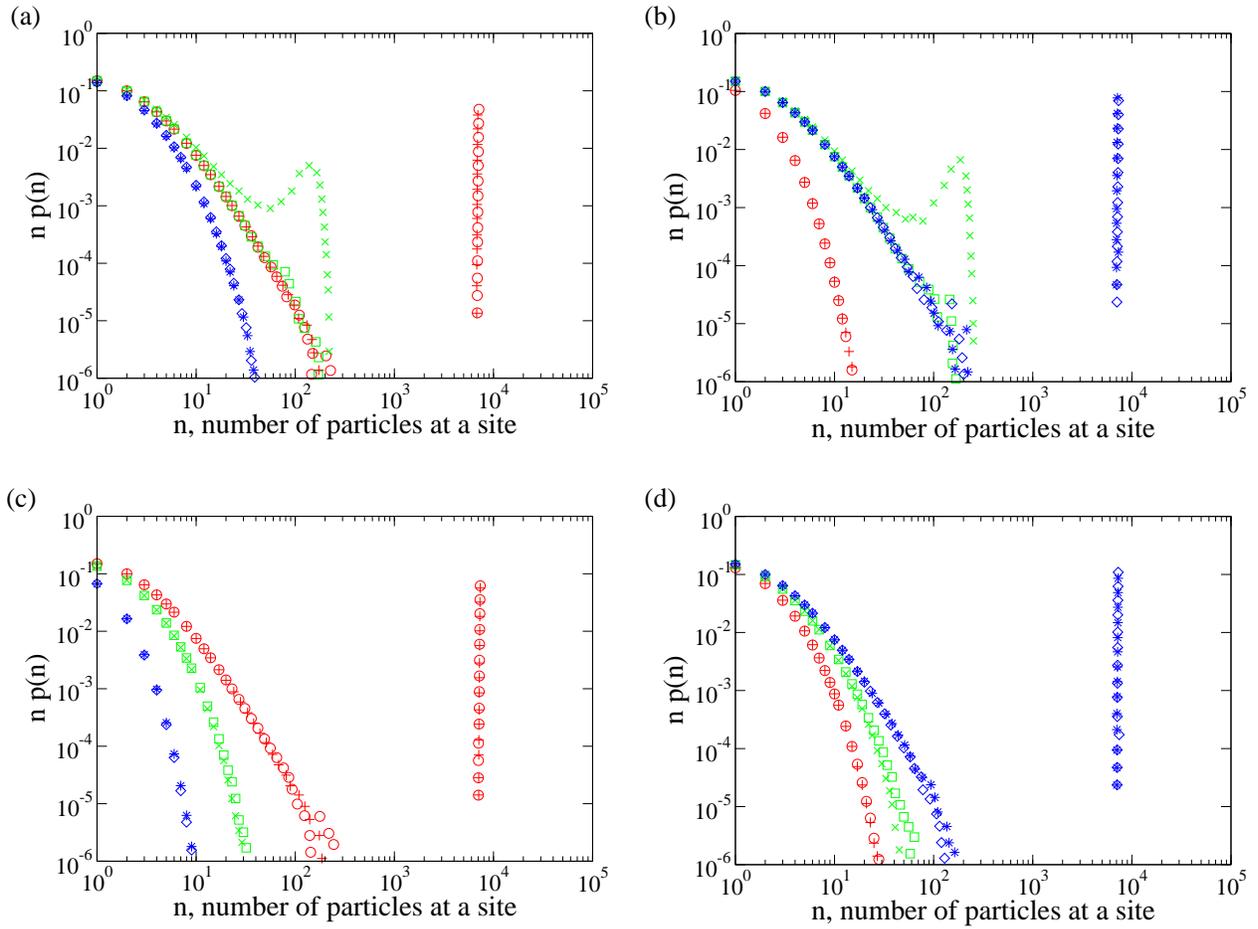}
\end{center}
\caption{
%Distributions of the number of particles on a site 
Residence plots of the number of particles on a site
in the main
stretch, left branch and right branch for the system with branch choice
probability, $x$, dependent on the state of the left branch only (given by
the form (\ref{xleft})) compared with the same distributions from a system
with the branch choice probability fixed at the value predicted by theory.
The system had $L=1900$ sites with $L_{\ell}=300$ and $L_r=600$, $N_{\mathrm{%
tot}}=8000$ particles and repulsion parameter $Q=20$. The distributions are
shown for $p$ and $q$ values which fit in the various phases: (a) LM phase ($%
p=0.25$, $q=0.85$), (b) LR phase ($p=0.1$, $q=0.5$), (c) M phase ($p=0.9$, $%
q=0.65$), (d) R phase ($p=0.58$, $q=0.25$).  In all cases the distributions
are given for the fixed-$x$ system in the main stretch (${\color{red} \circ}$), the
left branch (${\color{green} \square}$) and the right branch 
(${\color{blue} \diamond}$) and for the 
varying-$x$ system in the main stretch (${\color{red}+}$), the left branch 
(${\color{green}\times}$) and 
the right branch (${\color{blue}\ast}$).
From the theory it is expected
that the distributions from the system with $x$ fixed and the system with $x$
varying should match, with the exception of the condensate region. 
This is seen in (a) and (b) where in the fixed-$x$ system the condensates are
on the main stretch and right branch respectively and in the varying-$x$
system small amounts of these are shared with a condensate on the left branch.
%In the
%fixed-x system the condensate should be only on one site, but in the
%varying-x system the condensate can be split onto two sites. 
In general the
agreement is very good. 
%The apparent disagreement for the left branch with a
%small condensate is consistent with finite size effects for ring-lattice
%systems with particle densities only just above critical. The slight
%discrepancy in the left branches for (d) in the R phase is thought to be due
%to enhanced repulsion from the left branch due to the varying $x$ which
%suppress positive fluctuations.
}
\label{fxcomplbr}
\end{figure}

\begin{figure}[tbp]
\begin{center}
\includegraphics[width=\textwidth,angle=0]{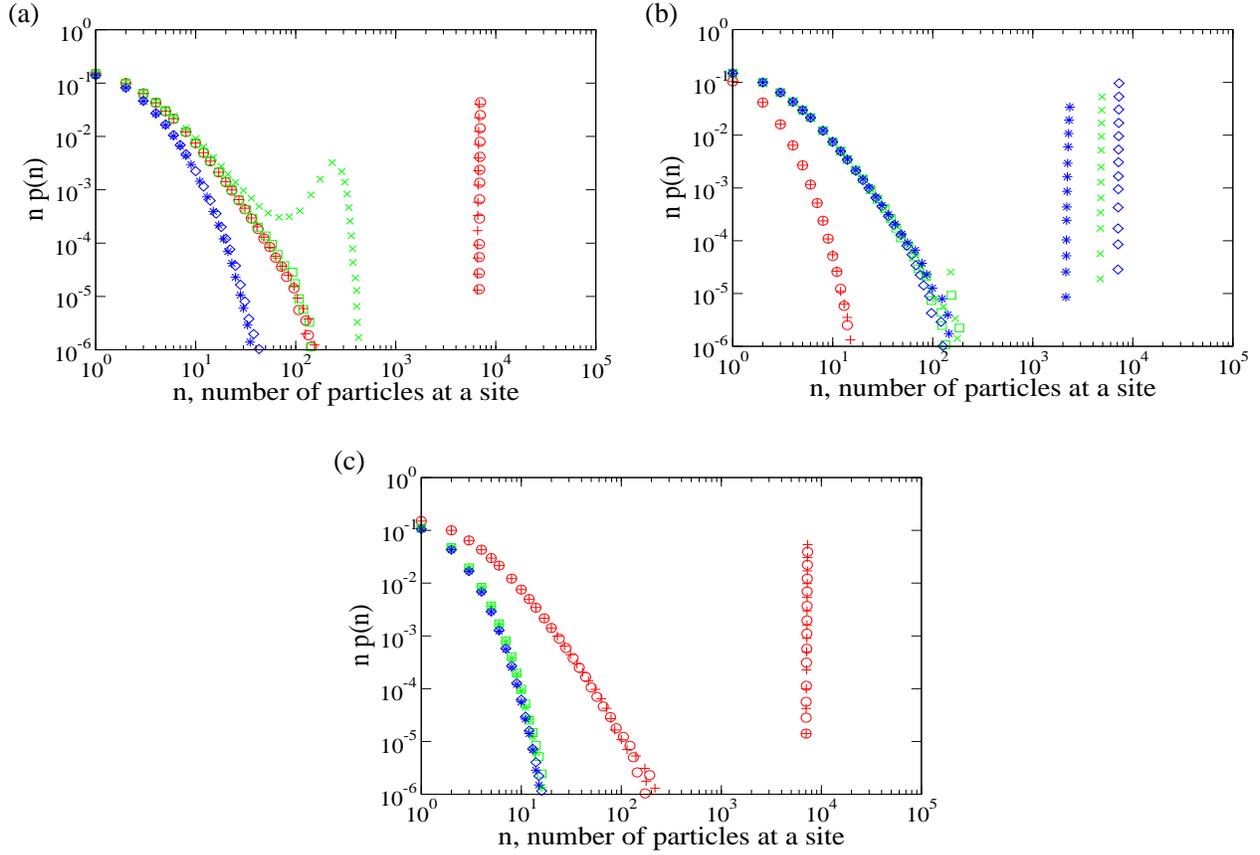}
\end{center}
\caption{%Distributions 
Residence plots 
of the number of particles on a site in the main
stretch, left branch and right branch for the system with branch choice
probability, $x$, dependent on the state of both branches (given by the form
(\ref{xrep})) compared with the same distributions for a system with $x$
fixed at the value predicted by theory. The system had $L=2000$ sites with $%
L_{\ell}=L_r=500$ and $N_{\mathrm{tot}}=8000$ particles. The distributions
are shown for $p$ and $q$ values which fit in the phases: (a) LM phase ($%
p=0.25$, $q=0.85$), (b) LR phase ($p=0.2$, $q=0.4$), (c) M phase ($p=0.75$, $%
q=0.85$). The RM phase is not represented as this is symmetric with the LM
phase. 
In all cases the distributions are given for the fixed-$x$ system in the main
stretch (${\color{red}\circ}$), the left branch 
(${\color{green}\square}$) and the right branch 
(${\color{blue}\diamond}$)
and the varying-$x$ system in the main stretch (${\color{red}+}$), the left branch
(${\color{green}\times}$) and the right branch (${\color{blue}\ast}$).
From the theory it is expected that the distributions from the system
with fixed $x$ and the system with varying $x$ should agree, except for the
part of the distribution which describes the condensate. For this system the
agreement is very good. 
%The only apparent discrepancy is for the left branch
%distribution in the LM phase (a). However, this is consistent with finite
%size effects in systems where the density is only a small amount above the
%critical density.
}
\label{fxcomprep}
\end{figure}

We also considered a more sensitive test for our mean-field theory. Since
the predictions from this approach are based on the solution of a system
with \emph{fixed }$x$, we performed simulations on such systems - with $x$ 
\emph{fixed} at the value expected by theory. In particular, we studied
distributions for the occupation of a site on each of the three sections
(M,L,R) for both these fixed-$x$ systems and the original model. Comparisons
are shown in Figures~\ref{fxcomplbr} and \ref{fxcomprep}. To be consistent
with the theory, these distributions should match everywhere \emph{except} 
in those parts which represent a condensate. Specifically, for the 
varying-x system at high enough densities, the single condensate characteristic 
of the fixed-$x$ system can be split into two
residing on two different sections of the lattice. In general, 
the agreement was observed to be excellent. There was some apparent discrepancy
when a ``small'' condensate was present, but such differences are consistent
with finite-size effects observed in the standard ZRP at densities above,
but close to, criticality.

\begin{figure}[tbp]
\begin{center}
\includegraphics[width=\textwidth,angle=0]{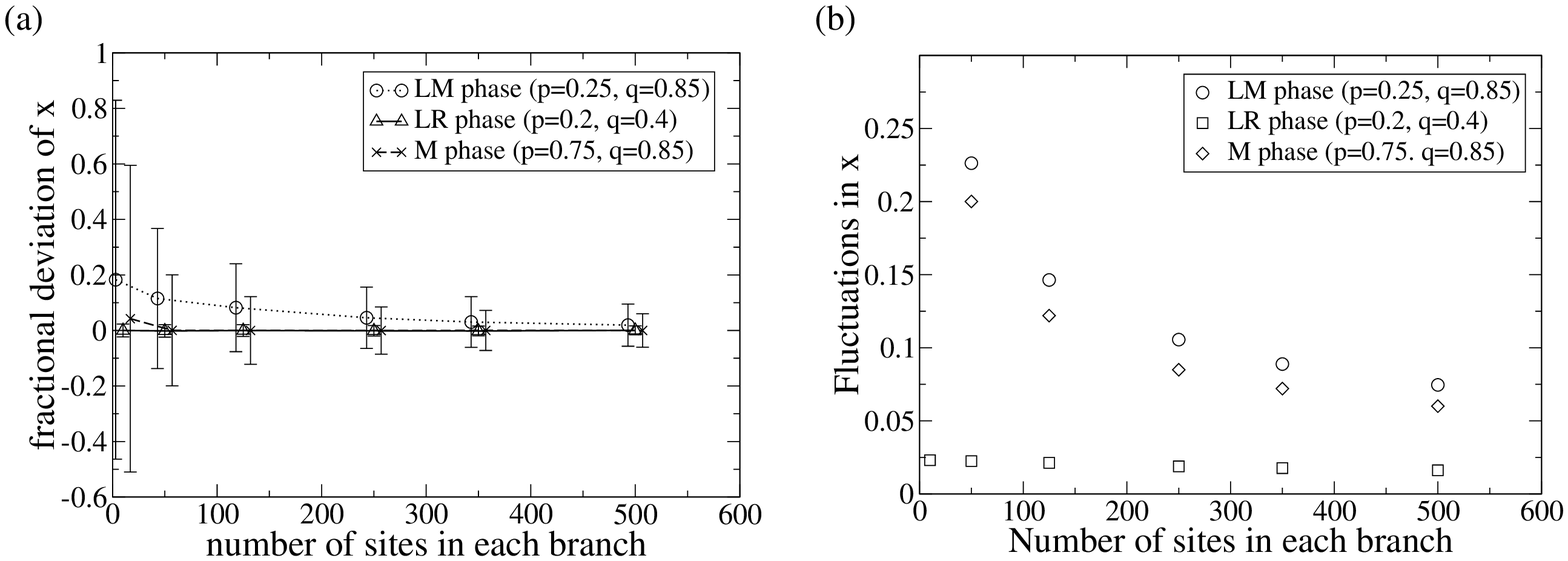}
\end{center}
\caption{Behaviour of the system with branch choice probability, $x$,
dependent on left and right branches (with the form given by (\ref{xrep})
with varying, but equal, numbers of sites in the branches. The system has $%
L_m=1000$ sites in the main stretch and a fixed global density of $4$
particles per site. (a) The fractional deviation of the branch choice
probability, $x$, from the value predicted by the mean-field theory. 
%For the
%LR and M phases the agreement is excellent for all branch lengths and the
%error becomes smaller as the length increases. For the LM phase there is a
%clear worsening of the agreement as the length decreases. This is probably
%due to the condensate on the left branch being rather small and as such
%being somewhat unstable and subject to large negative fluctuations. 
(b)
Fluctuations of the branch choice probability, $x$, on the scale of the
mean. 
%For the LR phase, fluctuations are small and show no discernible trend
%with system size. For the LM and M phases the fluctuations show a distinct
%increase as the length of the branches is decreased. This behaviour is most
%likely due to the system beginning to feel discrete effects due to the short
%lengths of the branches. The LR phase is unaffected because there are many
%particles on each branch in this phase.
}
\label{xdevfluc}
\end{figure}

We also studied how the quality of the agreement between the theory and
simulations depends on the lengths of the branches, focusing only on the
first feedback mechanism - $x$ given by (\ref{xrep}). We chose $L_{\ell}
=L_r$ in the range of 10 to 500 sites and ran the system in the LM, LR and M
phases. The global density of particles was fixed. We measured both the
fractional deviation of $x$ from its expected value and the fluctuations 
in $x$ on the scale of the mean, see Figure~\ref{xdevfluc}. Generally, the 
agreement worsens as the branches are shortened. However, this general trend is 
\emph{not} observed for the LR phase. One possible difference is the
following. For the LM and M phases, at least one of the branches will 
have a low density of occupation, leading
to effects of discreteness. In very short branches and with low densities,
there are typically only a few particles on the sites, so that changes in
occupation happen through comparatively large jumps. Also, the fact that the LM
phase deviates and fluctuates more than the M phase is probably due to the
fact that the condensate on the left branch is relatively small and so, more
susceptible to instability and collapse. For very small branch lengths,
some blurring of the phase boundaries was observed. Close to the boundaries
expected from theory, the states observed sometimes did not match with those
predicted. It is thought that these states may be metastable, as systems
artificially nucleated in the expected state remained there for the duration
of all runs.

Finally, we also studied numerically the validity of the theory at low
densities where no condensation occurs. Through simulations, we can deduce
the values of $x$ and $z$. These values are then inserted into  
(\ref{lbxconsistency}) and (\ref{densityeqn}) to see how well the latter are
satisfied. In all cases the agreement is very good. Apparently, the
grand-canonical analysis is quite successful for a self-consistent,
mean-field approach, despite suggestions that this type of analysis should
work well only for the condensed state.

\section{Conclusion}

\label{conclusion}

In this paper a generalization of a ZRP with limited long-range interactions
was studied. The basic model was that of a ring lattice with a section which
splits into two branches at a T-junction before rejoining. Apart from the
branch point (BP), we impose the standard ZRP rules for totally
asymmetric particle hopping. The rates are uniform within each branch, but
may be different for the three sections. For the particle at the BP, we must
assign probabilities to hop to each branch. If these are fixed parameters,
there exists an exact solution for the stationary state distribution. Here,
we focus on a simple generalization, i.e., that these probabilities depend
on the total occupation within the branches. Thus, there is a \emph{%
long-range interaction} between the particle at the BP and those in the
branches. We study this system both analytically and with extensive Monte
Carlo simulations. Our self-consistent mean-field theory is based on the
exact solution with fixed probabilities.

The long-range feedback mechanism has interesting consequences for the
system. For systems with sufficiently high densities and fixed branching
probabilities, a condensate forms only on one of the three sections. In other
words, there can be only four regions in a phase diagram (involving the overall
density, the relative hopping rates, and the branching probabilities). By
contrast, our model displays three \emph{additional} phases, associated with \emph{%
coexistence} of condensates on two of the three branches. As a side remark, we
note that coexistence of condensates on all three branches can only happen 
on a line rather than in an extended region in parameter space and is therefore
difficult to observe in simulations. As a result, the global phase diagram is
considerably richer.

In simulations, we often observe the condensate forming on the first site of
a section. This behaviour parallels the one displayed in a ZRP with \emph{%
open boundaries} \cite{EMS05}, in which particles are inserted into the left
boundary and extracted from the right boundary with constant rates. If the
insertion rate is low and the extraction rate is high, 
no condensate appears and much can be
understood through a treatment similar to the one used here: exploiting a
grand-canonical function with fugacities dependent on the boundary rates.
For high insertion and/or low extraction rates, condensation was seen on the end sites,
but these condensates tended to continue growing with time. In our branching
ZRP (with periodic boundary conditions), the formation of a condensate will
feed back into the equivalent of the insertion rate and so, a condensate with
a stable size can form. The success of the grand-canonical treatment in the
open boundary model may be related to the good agreement between the
mean-field theory and simulation results in our model.

Remarkably, this agreement continues to be quite good for very short (%
$\sim 10$ sites) branches. 
%\{\{\{\{Can a 1-site branch be solved
%analytically???\}\}\}\} 
This is somewhat surprising, since mean-field
theories are expected to be reliable only in the thermodynamic limit. We
suspect, however, that the effects due to the finite size of the \emph{main
stretch} may be more serious, and we intend to explore these effects in a
future publication. In any case, the good agreement we found provides hope
that the mean-field approach may be suitable for further generalizations,
such as more branches/loops and increasingly complex long-range feedbacks.
We believe that, with the promise of further surprises, such systems deserve
further investigations. In particular, we believe that there is considerable
potential for applying such models to a broad spectrum of physical systems.
In particular, there are many transport processes on networks where
individual ``agents'' make decisions on which route to take, based on the
existing state of the rest of the network. Examples include vehicular
traffic systems and data transport in computer networks. The model studied
here is perhaps the simplest of this class and we hope that it serves as a
springboard for complex generalized models and advances the understanding of
realistic systems.

\section{Acknowledgements}
This work was supported in part by a grant from the US National Science
Foundation DMR-0414122.

\end{document}